\documentclass[]{article}  
\usepackage{url}
\usepackage{graphicx}
\usepackage{amsfonts}
\usepackage{amssymb}
\usepackage{latexsym}

\newenvironment{pf}{\unskip{\bf Proof:}}{\unskip{\hfill $\Box$}}

\newcommand{\lemlab}[1]{\label{lemma:#1}}

\newcommand{\tablab}[1]{\label{tab:#1}}
\newcommand{\figlab}[1]{\label{fig:#1}}

\newcommand{\figref}[1]{\ref{fig:#1}}

\newtheorem{theorem}{Theorem}[section]
\newtheorem{lemma}[theorem]{Lemma}

\newtheorem{df}{Definition}[section]

{\catcode`\@=11
\gdef\setft#1#2#3{%
\def\@oddfoot{
{\setbox0=\hbox{#1}
\setbox1=\hbox{#3}
\ifdim\wd0>\wd1
\dimen0=\wd0
\box0\hfil#2\hfil\hbox to\dimen0{\hfil\hfil\box1}
\else \dimen0=\wd1
\hbox to\dimen0{\box0\hfil }\hfil#2\hfil\box1 \fi
}}} }

\def\complaint#1{}
\def\withcomplaints{
\newcounter{mycomplaints}
\def\complaint##1{\refstepcounter{mycomplaints}%
\ifhmode%
\unskip%
{\dimen1=\baselineskip \divide\dimen1 by 2 %
\raise\dimen1\llap{\tiny -\themycomplaints-}}\fi%
\marginpar{\tiny [\themycomplaints]: ##1}}%
}

\withcomplaints

\def\int{{\mathrm{int}}}

\begin{document}

\title{A Note on Objects Built From Bricks\\without Corners}

\author{%
Mirela Damian%
   \thanks{Department of Computer Science, Villanova University, Villanova,
PA 19085, USA.
   \protect\url{mirela.damian@villanova.edu}.}
\and
Joseph O'Rourke%
    \thanks{Department of Computer Science, Smith College, Northampton, MA
      01063, USA.
      \protect\url{orourke@cs.smith.edu}.
       Supported by NSF Distinguished Teaching Scholars award
       DUE-0123154.}
}
\maketitle

\begin{abstract}
We report a small advance on a question raised
by Robertson, Schweitzer, and Wagon
in~\cite{rsw-bo-02}.
They constructed a genus-13 polyhedron built from bricks without
corners, and asked whether every genus-0 such polyhedron
must have a corner.  
A \emph{brick} is a parallelopiped, and
a \emph{corner} is a brick of degree three or less
in the brick graph.
We describe a genus-3 polyhedron built from bricks with no corner,
narrowing the genus gap.
\end{abstract}

\section{Introduction}
Sibley and Wagon~\cite{sw-rpt3c-00}
proved that any collection of paralleograms glued whole-edge to
whole-edge must have at least one ``elbow'':  
a parallelogram with at most two neighbors.
This enabled them to prove that such tilings are
$3$-colorable.
The analogous question in 3D is~\cite{w-mrfch-02,do-op02-03}:
Must every object built from paraellopipeds (henceforth,
\emph{bricks}) have at least one corner, a brick with
at most three neighbors?
The bricks must be \emph{properly joined}:
each pair 
is either disjoint, or intersects either in a
single point,
a single whole edge of each, or a single whole face of each.
Two bricks in a collection are \emph{adjacent} if they share a single whole face.
Define the \emph{brick graph} of a collection of bricks to have
a node for each brick, and an arc for each pair of adjacent bricks.
The question is whether there must exist a node of degree $\le 3$
in the brick graph.
If so, $4$-colorability could be established.

The answer is {\sc no}: Robertson, Schweitzer, and Wagon
found a polyhedron with no corner.
This settled one question but raised another:  Might this be
true for a topological \emph{ball}, i.e., an object of genus~0
(their Question~1)?
Their example has a high genus; we show below its genus is $13$.
The main purpose of this note is to describe another example
that has genus~3.

\section{The Buttressed Octahedron}
Fig.~\figref{stan.oct} shows the $52$-brick
example from~\cite{rsw-bo-02}, in two views.
As shown, it has corners, but when each brick is partitioned
into eight sub-bricks from its center, it is an object
built from bricks with no corners.
\begin{figure}[htbp]
\centering
\includegraphics[width=0.95\linewidth]{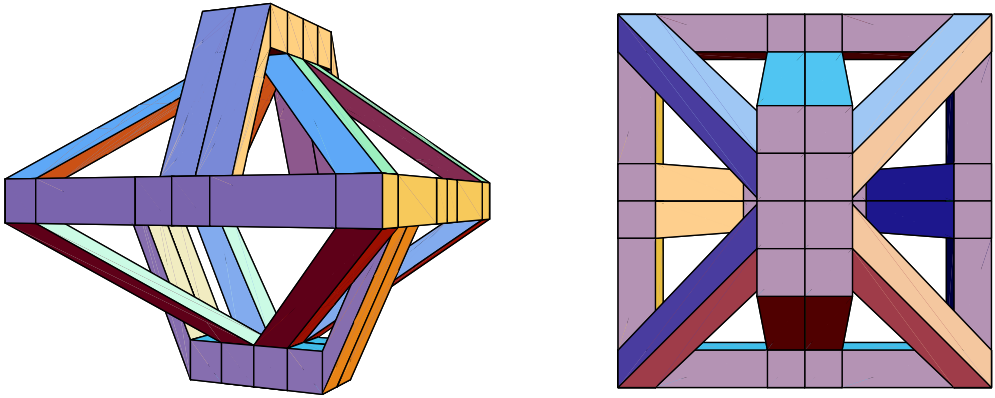}
\caption{An object with no corners (when refined).
[Fig.~1 from \protect\cite{rsw-bo-02}, by permission.]
}
\figlab{stan.oct}
\end{figure}

The authors did not report its genus.
We compute the genus from
the Euler characteristic $V-E+F=\chi$, which is
equal to $2 - 2g$, where $g$ is the genus.
This computation is performed on the unrefined object in the figure;
because $\chi$ is a topological invariant, 
it is unaltered by refinement/splitting.
The calculations are shown in Table~1. 

\begin{table}[htbp]
\begin{center}
\begin{tabular}{| l | c | c | c |}
	\hline
Pieces
	& $V$
	& $E$
	& $F$
	\\ \hline \hline

ring ($4$ quarters)
	& $4 \times 20 = 80$
	& $4 \times 40 = 160$
	& $4 \times 16 =64$
	\\ \hline
arch $(2)$
	& $2 \times 30 = 60$
	& $2 \times 66 = 132$
	& $2 \times 32 = 64$
	\\ \hline
buttress $(8)$
	& $0$
	& $8 \times 4 = 32$
	& $8 \times 4 = 32$
	\\ \hline \hline
sum
	& $140$
	& $324$
	& $160$
	\\ \hline

\end{tabular}
\caption{Euler characteristic calculations
$V-E+F=-24$.}
\end{center}
\tablab{Euler.Stan}
\end{table}

\begin{eqnarray*}
V-E+F & = & 140 - 324 + 160 \; = \; 2 - 2g\\
 g & = & 13
\end{eqnarray*}

\section{The ZZ-Object}
The overall design of our new example is shown in
Fig.~\figref{ZZ.int}.
It consists of two {\tt Z}-shaped paths
connecting four cubes.
Each of the long connectors has no corner when split lengthwise into four
bricks.  Similarly, the four cubes have no corners when split into eight cubes.
Prior to splitting, the object consists of only $10$ bricks.
However, as is evident from the figure, it is self-intersecting.
\begin{figure}[htbp]
\centering
\includegraphics[width=0.75\linewidth]{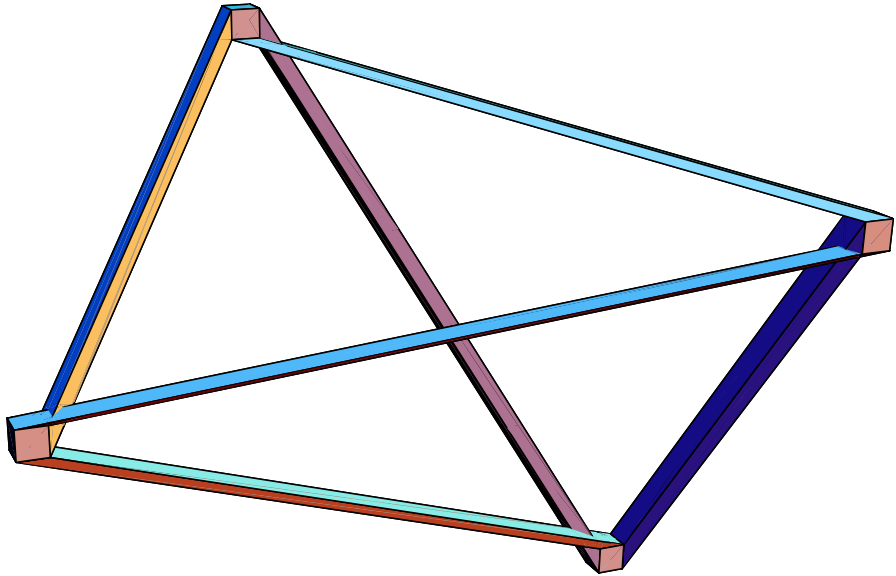}
\caption{A self-intersecting object with no corners (after refinement).}
\figlab{ZZ.int}
\end{figure}
In the figure, the centers of the four cubes are staggered 
at these coordinates:
\begin{eqnarray*}
& & ( 30, 40, 50 ) \\
& & ( 60, 10, 40 ) \\
& & ( 10, 20, 30 ) \\
& & ( 55, 30, 10 )
\end{eqnarray*}
The design of the object relies on this observation:
If it can be arranged that every brick has two opposite faces
covered by other bricks, then splitting will render it cornerless,
with every sub-brick with $\le 2$ exposed faces, and so of degree $\ge 4$.

The self-intersection can be removed by zig-zagging one of
the {\tt Z}s.
The resulting object of $14$ bricks is shown in 
Fig.~\figref{ZZ.noint}, in two views.
\begin{figure}[htbp]
\centering
\includegraphics[width=0.75\linewidth]{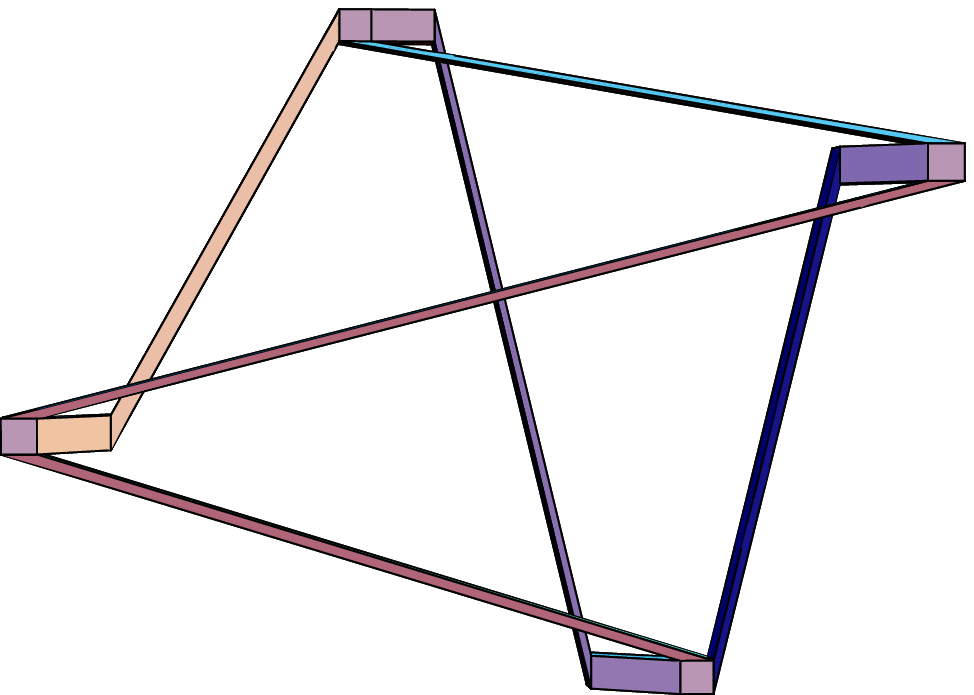}
\includegraphics[width=0.75\linewidth]{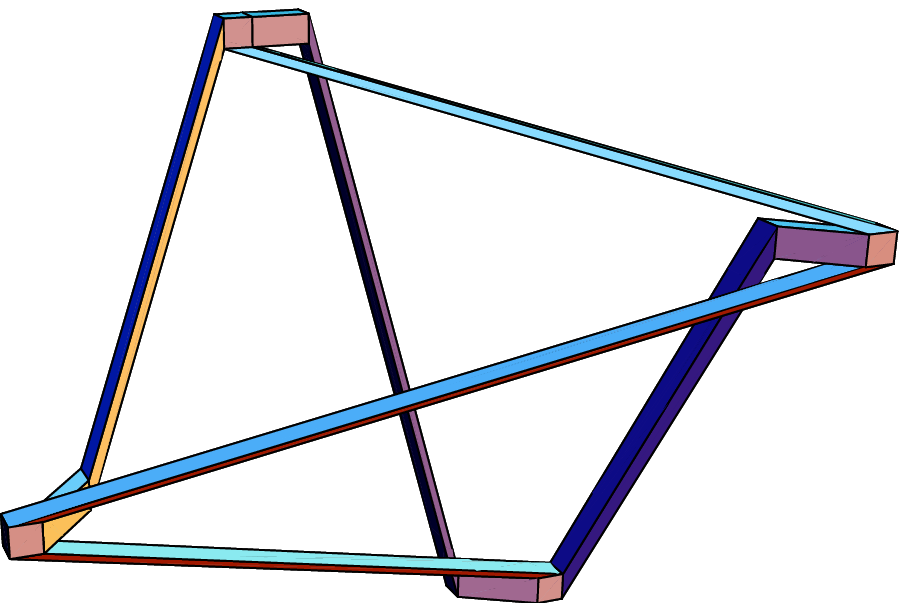}
\caption{A non-self-intersecting object of genus $3$ with no corners (after refinement).}
\figlab{ZZ.noint}
\end{figure}

We again compute the Euler characteristic $\chi$.
Because of its topological invariance, we compute it for
the simpler object in Fig.~\figref{ZZ.int}, which
has the same
genus as the object in Fig.~\figref{ZZ.noint} .  We count all vertices as part of the cubes.
All $12$ edges of every cube are exposed, but only $3$ faces of each.
Each of the  $6$ connecting bricks contributes $4$ edges and $4$ faces.
The calculations are shown in Table~2. 

\begin{table}[htbp]
\begin{center}
\begin{tabular}{| l | c | c | c |}
	\hline
Pieces
	& $V$
	& $E$
	& $F$
	\\ \hline \hline

cubes $(4)$
	& $4 \times 8 = 32$
	& $4 \times 12 = 48$
	& $4 \times 3 =12$
	\\ \hline
connectors $(6)$
	& $0$
	& $6 \times 4 = 24$
	& $6 \times 4 = 24$
	\\ \hline \hline
sum
	& $32$
	& $72$
	& $36$
	\\ \hline

\end{tabular}
\caption{Euler characteristic calculations
$V-E+F=-4$.}
\end{center}
\tablab{Euler.ZZ}
\end{table}

\begin{eqnarray*}
V-E+F & = & 32 -72 + 36 \; = \; 2 - 2 g\\
 g & = & 3
\end{eqnarray*}

\section{Conclusion}
We have constructed a $14$-brick, genus-$3$ object, 
which when refined by splitting,
has no corners.
The question remains whether there exists an object 
without corners of smaller genus: $2$, $1$, or $0$.  
We conjecture that all brick objects of genus~$0$ must have a corner.
This gains some support from the work in~\cite{go-cobb-03}
which shows that one class of topological balls always has
at least four corners. 

\paragraph{Acknowledgements.}
We thank
Sasha Berkoff,
Asten Buckles,
Jessie McCartney,
and Shawna King for assisting in the preparation of Figs.~2 and~3.

\bibliographystyle{alpha}
\bibliography{/home1/orourke/bib/geom/geom}

\end{document}